\documentclass[twoside,fleqn]{article}
\usepackage{epsfig}
\usepackage{espcrc2}

\def\lsim{\raise0.3ex\hbox{$<$\kern-0.75em\raise-1.1ex\hbox{$\sim$}}}
\def\Chi{\raise0.4ex\hbox{$\chi$}}

\def\Tr{\rm{tr}}

\newcommand{\beqn} {\begin{equation}}
\newcommand{\eqn} {\end{equation}}

\title{String breaking in Lattice QCD \thanks{
      Poster presented at the XVI International Symposium on
      Lattice Field Theory, Boulder, CO, July 13-18, 1998.}}

\author{E. Laermann\address{Fakult\"at f\"ur Physik,
        Universit\"at Bielefeld,
        P.O. Box 100 131, 33501 Bielefeld, Germany} 
        with
        C. DeTar\address{Department of Physics, University of Utah,
        Salt Lake City, UT 84112, USA}, 
        O. Kaczmarek$^{\rm a}$ and F. Karsch$^{\rm a}$
        }

\begin{document}

\begin{abstract}
The separation of a heavy quark and antiquark pair leads to the
formation of a tube of flux, or string, which should break in the presence
of light quark-antiquark pairs. This expected zero temperature
phenomenon has proven elusive in simulations of lattice QCD.
We present simulation results that show that the string does
break in the confining phase at nonzero temperature. 
\end{abstract}

\maketitle

In the absence of light quarks the heavy quark-antiquark
potential is known quite accurately from numerical simulations
of lattice quantum chromodynamics \cite{quenchedpot}. 
At large separation $R$,
the potential rises linearly, as expected in a confining
theory. 
In the presence of light quarks it is expected that the string
between the heavy quark-antiquark pair breaks at large distance.
All the existing lattice data at zero temperature 
\cite{fullpot1,fullpot2} agree in that they do not
show any indication of string breaking which would be
signalled by a tendency 
of the potential to level off at
large distances. The distances covered so far extend 
up to $R \lsim 2$ fm while 
it has been proposed that the dissociation threshold
would 
be reached at separations somewhere 
between 1.5 and 1.8 fm \cite{fullpot2,diss}.


We have simulated QCD with two light flavours 
of staggered dynamical 
quarks on lattices of size $16^3 \times 4$
(new work) and $12^3 \times 6$
(configurations from Ref.~\cite{eos6})
at fixed values for the quark mass of
$m_q/T = 0.15$ and $0.075$ respectively. The couplings were
chosen 
to cover temperatures $T$ below the critical temperature
$T_c$ in the range of approximately
$0.7 T_c < T < T_c$. 
The (temperature-dependent) 
heavy quark potential $V(R,T)$ was extracted
from Polyakov loop correlations
\beqn
\langle L({\vec 0}) L^\dagger ({\vec R}) \rangle = c \, 
\exp\{ - V(|{\vec R}|,T)/T\}
\eqn
where 
$
L({\vec x})= \frac{1}{3} \Tr \prod_{\tau = 0}^{N_\tau - 1}
U_0({\vec x},\tau)
$
denotes the Polyakov loop at spatial coordinates ${\vec x}$.
In the limit $R \rightarrow \infty$ the correlation function
should approach the cluster value
$|\langle L(0)\rangle |^2$ which vanishes if the potential
is rising at large distances (confinement) 
and which acquires a small but
finite value if the string breaks.

\begin{figure}[htb]
 \epsfig{bbllx=115,bblly=340,bburx=540,bbury=735,clip=,
         file=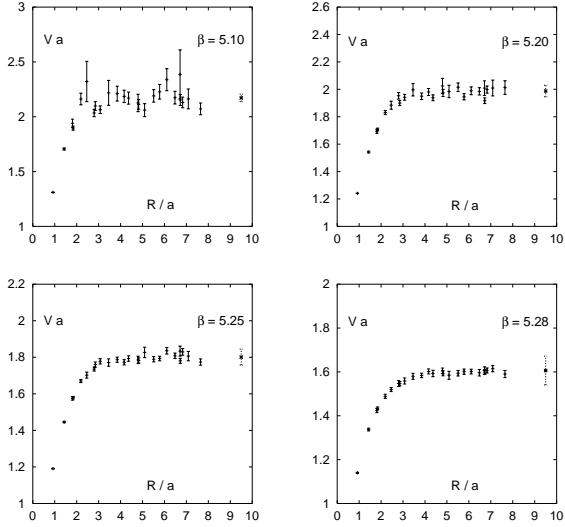,width=75mm}
\vspace{-10mm}
\caption{The potentials in lattice units at the $\beta$ values
         analyzed for $N_\tau=4$. The right-most data points 
         plotted at $R/a = 9.5$ and denoted by stars 
         are the infinite distance
         cluster values $-T {\rm ln} |\langle L \rangle |^2$.
         }
\label{fig:nt4}
\end{figure}

\begin{figure}[htb]
 \epsfig{bbllx=110,bblly=280,bburx=555,bbury=710,
         file=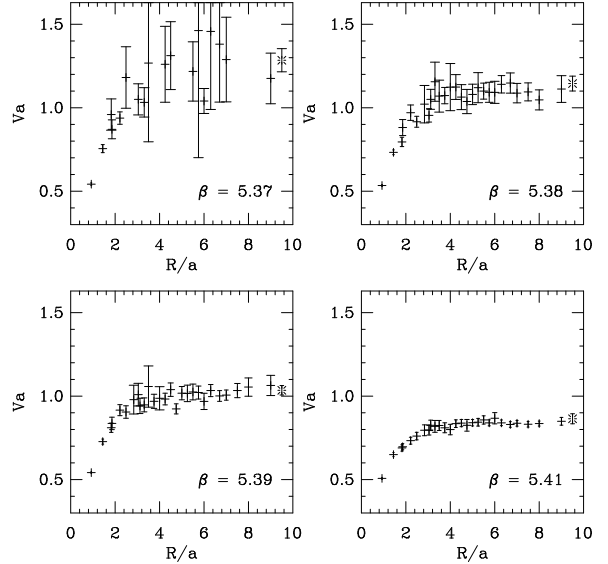,width=75mm}
\vspace{-10mm}
\caption{Same as Figure~1 except for $N_\tau=6$. 
         }
\label{fig:nt6}
\end{figure}

In Figures \ref{fig:nt4} and \ref{fig:nt6} our
data for the potential are presented in lattice units 
at the values of $\beta$ analyzed.
The critical couplings $\beta_c$ have been
determined as $5.306$ for $N_\tau = 4$ and $5.415$ 
for $N_\tau = 6$ respectively. The Polyakov loop correlations
have been computed not only for on-axis separations but
also for a couple of off-axis distance vectors ${\vec R}$.
Rotational invariance is reasonably well recovered if one
uses the lattice Coulomb behaviour to determine the quark-antiquark
separation, $|{\vec R}| = 1/G_{\rm lat}({\vec R})$.

The data in Figures \ref{fig:nt4} and \ref{fig:nt6} quite
clearly show a flattening of the potential at lattice
distances of about 3 to 4 lattice spacings, depending
on $\beta$.
This confirms earlier results \cite{Wien} obtained on smaller lattices
of size $8^3 \times 4$.
Moreover, the height of the potential at these distances is
in nice agreement already with the infinite distance cluster value,
shown as the right-most data point in each of the plots.

In order to obtain a rough estimate of the corresponding
temperatures in units of the critical temperature we applied
the following procedure: at the given $\beta$
and $m_q a$ values an 
interpolation formula \cite{MILCinter}
was utilized to estimate the vector meson mass $m_V a$ in lattice
units as well as the ratio of pseudoscalar to vector meson mass,
$m_{PS}/m_V$. By means of a phenomenological formula which
interpolates between the (experimentally measured) 
$\rho$ and $K^*$ mass as function
of the ratio $m_{PS}/m_V$, 
$
m_V = 756 {\rm MeV} + 450 {\rm MeV} \times \left( m_{PS}/m_V
\right)^2,
$
a physical value for $m_V$ can be obtained.
These numbers are then used to estimate the ratios of the lattice
spacing at the various $\beta$. 

Finally, in order to facilitate a comparison 
of the $N_\tau = 4$ and $6$ results with each other and
with quenched data, the absolute scale was determined from a 
conventional Wilson loop measurement
of the string tension at zero temperature at the critical
$\beta_c$ values. The Wilson loops did not show
string breaking at the separations which could be explored.
The results for the critical temperature in units of the
string tension are obtained as
$T_c/\sqrt{\sigma} = 0.436(8)$ 
for $N_\tau = 4$ and 
$T_c/\sqrt{\sigma} = 0.462(9)$ 
for $N_\tau = 6$ \cite{MILCsigma}. 

\begin{figure}[htb]
 \epsfig{file=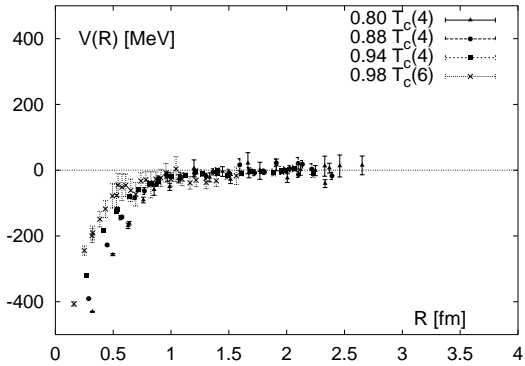,width=75mm}
\vspace{-14mm}
\caption{The potential in physical units at various temperatures.
         The results are from lattices with $N_\tau = 4$ and $6$, 
         indicated by the number in brackets. The data has been
         normalized to the cluster value.
}
\label{fig:norm}
\end{figure}

In Figure \ref{fig:norm} we show the potential in the
presence of dynamical quarks in physical units. 
The potential is flat within
the error bars at distances larger than about 1 fm. It also
seems that the turn-over point is slightly $T$ dependent,
becoming smaller with increasing temperature. 

\begin{figure}[t]
 \epsfig{file=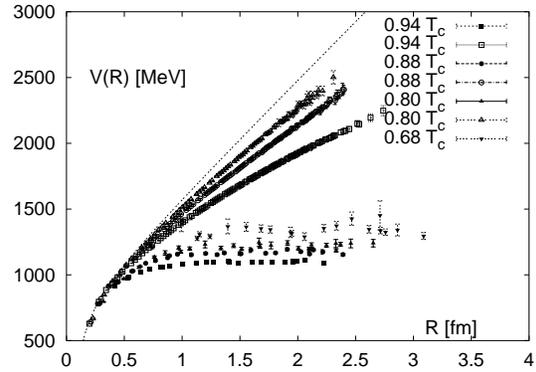,width=75mm}
\vspace{-14mm}
\caption{The potential in physical units at various temperatures.
         Compared are quenched (open symbols) and 
         full (filled symbols) QCD potentials at the same 
         temperature. The dashed line is the zero temperature
         quenched potential. 
         The data has slightly been shifted as to
         agree at distances around 0.3 fm.
}
\label{fig:phys}
\end{figure}

In Figure~\ref{fig:phys} quenched and full QCD potentials
are compared. The quenched data
has been taken from \cite{Olaf} and was obtained in the same
way, i.e. computed from Polyakov loop correlations. 
Figure~\ref{fig:phys} contains, for further
comparison, the dashed line denoting
$-\pi/(12 R) + (420 {\rm MeV})^2 R$ 
which gives a good description of
the zero temperature quenched potential.
Note that the finite temperature
quenched potential is rising with distance $R$ but the slope 
decreases with temperature, i.e. the (quenched)
string tension is temperature 
dependent and becomes smaller closer to the critical $T_c$. 
Again, the comparison with quenched potentials at the same
temperature demonstrates that
the potential in the presence of dynamical quarks becomes 
flat within the error bars at distances of about 1 fm.
From Figure~\ref{fig:phys} we conclude that the observed
string breaking, albeit at finite temperature, is an effect
caused by the presence of dynamical fermions. 

We have seen that string breaking is relatively easy to observe in the
Polyakov loop correlation, while it is difficult to detect through the
conventional Wilson loop observable.  Why is this so?  The Wilson loop
observable creates a static quark-antiquark pair together with a flux
tube joining them.  In the presence of such a static pair at large
$R$, we expect the correct ground state of the Hamiltonian to consist
of two isolated heavy-light mesons, however.  Such a state with an
extra light dynamical quark pair has poor overlap with the flux-tube
state, so it is presumably revealed only after evolution to a very
large temporal separation.  
An improved Wilson-loop-style determination of the heavy
quark potential in full QCD would employ a variational superposition
of the flux-tube and two-heavy-meson states \cite{drummond,Knechtli}. The
Polyakov loop approach, on the other hand, although limited in
practical application to temperatures close to or above $T_c$, 
builds in no
prejudices about the structure of the static-pair ground state wave
function.  Screening from light quarks in the thermal ensemble occurs
readily.


\begin{thebibliography}{99}
\bibitem{quenchedpot} 
         G. Bali and K. Schilling, Phys. Rev. D46 (1992) 2636, D47
            (1993) 661;
         S.P. Booth et al. (UKQCD Coll.), Phys. Lett. B294 (1992) 385;
         Y. Iwasaki et al., Phys. Rev. D56 (1997) 151;
         B. Beinlich et al., {\tt hep-lat/9707023};
         R.G. Edwards, U.M. Heller and T.R. Klassen, Nucl. Phys. B517
              (1998) 377
\bibitem{fullpot1} 
         K.D. Born et al., Phys. Lett. B329 (1994) 325;
         U.M. Heller et al., Phys. Lett. B335 (1994) 71;
         U. Gl\"assner et al. (SESAM Coll.), Phys. Lett. B383
            (1996) 98;
         C. Bernard et al. (MILC Coll.), Phys. Rev. D56 (1997) 5584;
         S. Aoki et al. (CP-PACS Coll.), Nucl. Phys. B(Proc. Suppl.)63A-C
            (1998) 221.
\bibitem{fullpot2} 
         M. Talevi et al. (UKQCD Coll.), Nucl. Phys. B(Proc. Suppl.)63A-C
            (1998) 227.
\bibitem{diss} C.Alexandrou et al., Nucl.Phys.B414(94)815.
\bibitem{eos6}
         T.~Blum (for the MILC Collaboration), 
         Nucl.\ Phys.\ B (Proc.\ Suppl.) 47 (1996) 503; 
         C.~Bernard {et al.} (MILC) Phys.\ Rev.\ D55 (1997) 6861.
\bibitem{Wien} 
         W.~Sakuler {et al.} Phys.\ Lett.\ B276 (1992) 155;
         W.~B\"urger et al., Phys.\ Rev.\ D47 (1993) 3034.
\bibitem{MILCinter} C. Bernard et al. (MILC Coll.), Phys. Rev. D54 (1996) 4585.
\bibitem{MILCsigma} C. Bernard et al. (MILC Coll.), Phys. Rev. D56 (1997) 5584.
\bibitem{Olaf} O.Kaczmarek, Diploma thesis, Bielefeld 1998.
\bibitem{drummond} 
        I.~Drummond, {\tt hep-lat 9805012}.
\bibitem{Knechtli}
        O. Philipsen and H. Wittig, {\tt hep-lat 9807020};
        F. Knechtli and R. Sommer, {\tt hep-lat 9807022}.
\end{thebibliography}
\end{document}